\documentclass[9pt, twocolumn]{article}
\usepackage{geometry}                
\usepackage{graphicx}
\usepackage{authblk}
\usepackage{amssymb,amsmath}
\usepackage{epstopdf}
\usepackage{enumerate}
\usepackage{url}
\usepackage{sectsty}

\sectionfont{\large}

\title{How ``Quantum'' is the D-Wave Machine?}
\author[*]{Seung Woo Shin}
\author[$\dag$]{Graeme Smith}
\author[$\dag$]{John A.~Smolin}
\author[*]{Umesh Vazirani}
\affil[*]{\textit{Computer Science division, UC Berkeley, USA.}}
\affil[$\dag$]{\textit{IBM T.J.~Watson Research Center, Yorktown Heights, NY 10598, USA.}}
\date{}                                           

\begin{document}
\maketitle
{\textbf{\textit{ Recently there has been intense interest in claims about the performance of the D-Wave machine. 
In this paper, we outline a simple classical model, and show that it achieves excellent correlation 
with published input-output behavior of the D-Wave One machine on 108 qubits. While raising questions
about ``how quantum'' the D-Wave machine is, the new model also provides additional algorithmic insights into the nature
of the native computational problem solved by the D-Wave machine.}}}

\smallskip
In a future world of quantum devices, it will become increasingly important to test that these devices behave according to specification. While this is clearly a central issue in the context of quantum cryptography~\cite{BHK05, VV14} and certified random number generators~\cite{PAM10, VV12}, it is also quite fundamental in the context of testing whether a claimed quantum computer is really quantum~\cite{ABE10,RUV13,FK12,BFKW13}. Recently, this last issue has featured prominently in the context of the D-Wave machine~\cite{16qubit, 8qubit, Boixo1, Scott, dwave2, RWJ14}, amidst questions about to what extent it is ``truly quantum'' and whether it provides speedups over classical computers. 

Of course, whether something is ``truly quantum'' depends upon the scale or level of abstraction. Thus a laptop computer is certainly not quantum even though quantum mechanics is essential to the design and description of its central computing element, the transistor. By contrast, a minimal requirement for even a special purpose quantum computer is that it exhibits large-scale quantum behavior. For all practical purposes this means large-scale entanglement. 

In this paper, we investigate whether the 108-qubit D-Wave machine exhibits large-scale quantum behavior, by analyzing recently released experimental data of D-Wave One \cite{Comment}.  We show that there is a natural classical model that demonstrates excellent correlation with the input-output behavior of the D-Wave machine as recorded in this data. 

Our model is simple. Qubits are modeled as classical magnets coupled through nearest-neighbor interaction and subject to an external magnetic field. The magnitude of these interactions is borrowed directly from the D-Wave data. The finite temperature of the device is modeled by applying the Metropolis rule to randomly ``kick" each magnet at each step. The success of such a simple model at explaining the existing data on the large scale input-output behavior of the D-Wave machine naturally raises questions about whether the D-Wave machine, at a suitable level of abstraction, is better described by a  classical model. 

At a more abstract level, our model may be regarded as carrying out simulated annealing on 2D vectors rather than bits (or classical spins), but in the presence of an external transverse field. Why then does our model correlate so well with the D-Wave machine while simulated annealing does not? The key to understanding this is a new algorithmic insight: while neither feature by itself (2D vectors and transverse field) changes the qualitative behavior of the algorithm, surprisingly both features make the algorithm behave very differently from classical simulated annealing and more like belief propagation. 

A deeper understanding of this new algorithm sheds new light on the nature of quantum annealing, the quantum algorithm that the D-Wave machine seeks to implement. Quantum annealing is the finite temperature implementation of adiabatic quantum optimization. The hope for speedup by adiabatic quantum optimization (and by extension for quantum annealing) lies in the possibility of tunneling through local optima. Theoretical studies have only demonstrated tunneling in extremely special circumstances (see~\cite{R04}), and a sequence of papers prove that the quantum adiabatic algorithm gets stuck in local optima resulting in exponential worst case behavior on large classes of instances~\cite{vDMV02, vDV03, R04}. Any optimism about the prospects of speedup by quantum annealing rely on the hope that the energy landscapes for practical instances of optimization problems have some special structure that facilitates tunneling. This is why there was great interest in the results of Boixo et al.~\cite{Boixo1} demonstrating that the input-output behavior of the D-Wave machine correlated strongly with simulated quantum annealing but not with classical simulated annealing. These results appeared to both provide evidence in favor of tunneling as well as quantum coherence in the D-Wave machine, despite the short decoherence times for individual qubits. Our classical model demonstrates that tunneling may not be necessary to explain this input-output behavior of D-Wave.

The new model also provides interesting algorithmic insights into the native computational problem solved by the D-Wave machine, which is to find the ground state
of a classical Ising spin glass on a certain interaction graph. The interaction graph for the D-Wave machine is the so-called Chimera graph, which consists of 16 clusters,
each consisting of 8 qubits. Experiments with our model indicate that since the clusters are highly connected, they effectively act as supernodes, thereby reducing 
the effective problem size from 108 to 16. This insight also suggests that as the number of qubits is scaled up, and the number of clusters increases from 16 to 64 and 
beyond, combinatorial explosion will start kicking in and the D-Wave machine will be ineffective at finding ground states. This insight is consistent with recent results
on the 512 qubit D-Wave II machine~\cite{RWJ14}.

In view of these results, one interesting direction to pursue is to ask whether there are regimes consisting of other classes of inputs on which the D-Wave machine exhibits ``truly quantum" behavior, in particular behaving differently from the model presented in this paper. Of course, our model is quite rudimentary, and makes no attempt to model details of the D-Wave machine such as errors in control of external fields and interaction strengths. For example, a class of instances has been proposed in a recent paper~\cite{Vinci}, but as we explain in a recent note~\cite{Vinci_response} a simple modeling of control errors reproduces the input-output behavior reported in the experiments in~\cite{Vinci}. More generally, establishing that a phenomenon is truly ``quantum" at a large scale is extremely challenging, since it involves ruling out all possible classical explanations. While this may not be practically feasible, it is difficult to overemphasize the importance of carefully ruling out a range of classical models. In particular, the note~\cite{Vinci_response} demonstrates the value of carefully considering elaborations of the rather rudimentary model presented in this paper while investigating how well it matches the behavior of a complex machine like D-Wave. 

\section*{The D-Wave Architecture and Tunneling}
The D-Wave architecture is a special purpose computer that is designed to solve a particular optimization problem, namely finding the ground state of a classical Ising spin glass. The classical Ising spin glass involves a set of classical spins interacting via nearest neighbor $z$-$z$ coupling. Formally the problem is specified by an interaction graph on $n$ vertices, together with interaction strengths $J_{ij}\in\{-1,1\}$ for each edge $\{i,j\}$ in the graph.\footnote{Here, we are presenting a simplifed version of the problem for the sake of clarity. In practice, $J_{ij}$ can be any real number between $-1$ and $1$ and spins can have local fields $h_i\in[-1,1]$, yielding the Hamiltonian $H=-\sum_{i<j} J_{ij}z_iz_j-\sum_i h_iz_i$. It is suggested in \cite{Boixo1} that the simplified version presented in the main text captures the hardest instances of this problem.} The ground state is a spin configuration with each spin value $z_i\in\{-1,1\}$ chosen to minimize the energy 
$H=-\sum_{i<j} J_{ij}z_iz_j$. 

\begin{figure}[!t]
\centering
\includegraphics[width=\columnwidth]{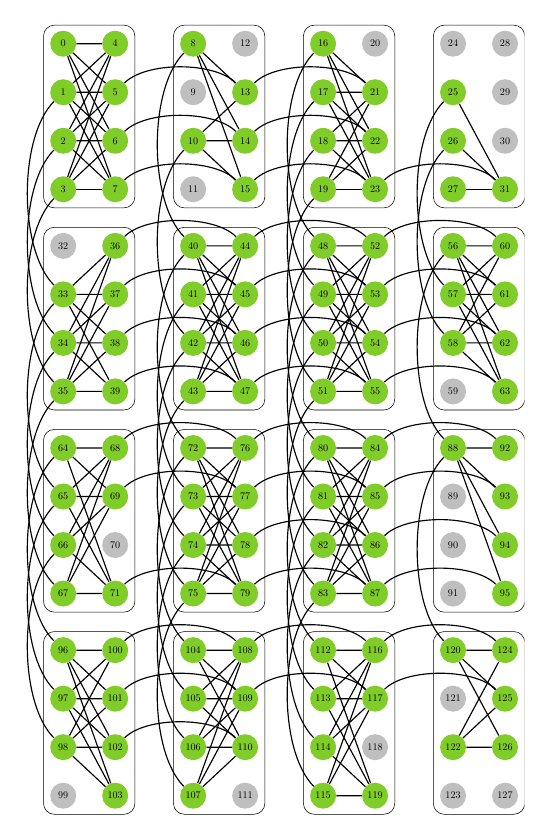}
\caption{The ``Chimera'' graph structure implemented by the D-Wave One. Grey dots represent the twenty non-working qubits in the D-Wave One. Figure is from \cite{Boixo1}.}\label{fig:chimera}
\end{figure}

In the current D-Wave architecture, the input graph is restricted to subgraphs of the so-called Chimera graph (Figure \ref{fig:chimera}). The Chimera graph may be described as a 2D lattice with each lattice point replaced by a supernode of $8$ vertices arranged as a complete bipartite graph $K_{4,4}$ (see Figure \ref{fig:chimera}). The left qubits in each supernode are coupled vertically in the 2D lattice and the right qubits horizontally. More specifically, each left qubit is coupled with the corresponding left qubit in the supernodes immediately above and below it, and each right qubit to the corresponding right qubits in supernodes immediately to the right and left. While the problem of finding the ground state of a classical Ising spin glass is NP-hard, it is known to admit a polynomial time approximation scheme when the input graph is a Chimera graph \cite{PTAS}.\footnote{Moreover, \cite{VChoi} makes the observation that the tree width of the Chimera graph scales as $\sqrt{n}$ rather than linearly and therefore the running time should also grow as $2^{O(\sqrt{n})}$ in the worst case.}

\begin{figure}[!t]
\centering
\includegraphics[width=0.8\columnwidth]{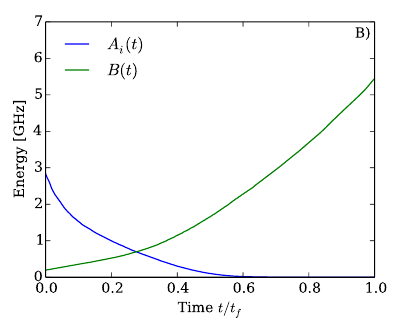}
\caption{The annealing schedule of the D-Wave One. Figure is from \cite{Boixo1}.}\label{fig:schedule}
\end{figure}

The D-Wave machine places a superconducting flux qubit at each node of the Chimera graph. The  time-dependent Hamiltonian of D-Wave machines is given by
\[
H(t) = -A(t)\sum_{1\leq i\leq n} \sigma_i^x -B(t)\sum_{i<j} J_{ij} \sigma_i^z \sigma_j^z
\]
where $A(t)$ and $B(t)$ represent the ``annealing schedule'' of the machine, which is shown in Figure \ref{fig:schedule}.
The process is carried out at a finite temperature.

If the entire process were carried out sufficiently slowly and at zero temperature, then it would be an implementation of adiabatic quantum optimization \cite{adiabatic}. Such a 
process is guaranteed to end up in the ground state of the target Hamiltonian $H=-\sum_{i<j} J_{ij}\sigma_i^z \sigma_j^z$ as long as the total annealing time scales as $\Omega(1/\Delta^2)$, where $\Delta$ is the minimum spectral gap of the time-dependent Hamiltonian where the minimum is taken over the annealing
schedule. The hope that this adiabatic optimization might lead to a speedup over classical algorithms lies in the possibility of quantum tunneling, whereby the algorithm tunnels through a barrier in the energy landscape rather than jumping over it due to thermal fluctuations (see \cite{R04} for the most positive example). However, it is known that in the worst case $\Delta$ scales exponentially in the problem size~\cite{vDMV02, vDV03, R04}, which means that adiabatic quantum optimization requires exponential running time in the worst case. There is also some numerical evidence \cite{AKR10} that certain optimization problems yield exponentially small gaps even on random instances. 

Quantum annealing can be thought of as a noisy, heuristic version of adiabatic quantum computing which is carried out at a finite temperature with an annealing time that does not respect the above spectral gap condition. 

\begin{figure}[!t]
\centering
\includegraphics[width=\columnwidth]{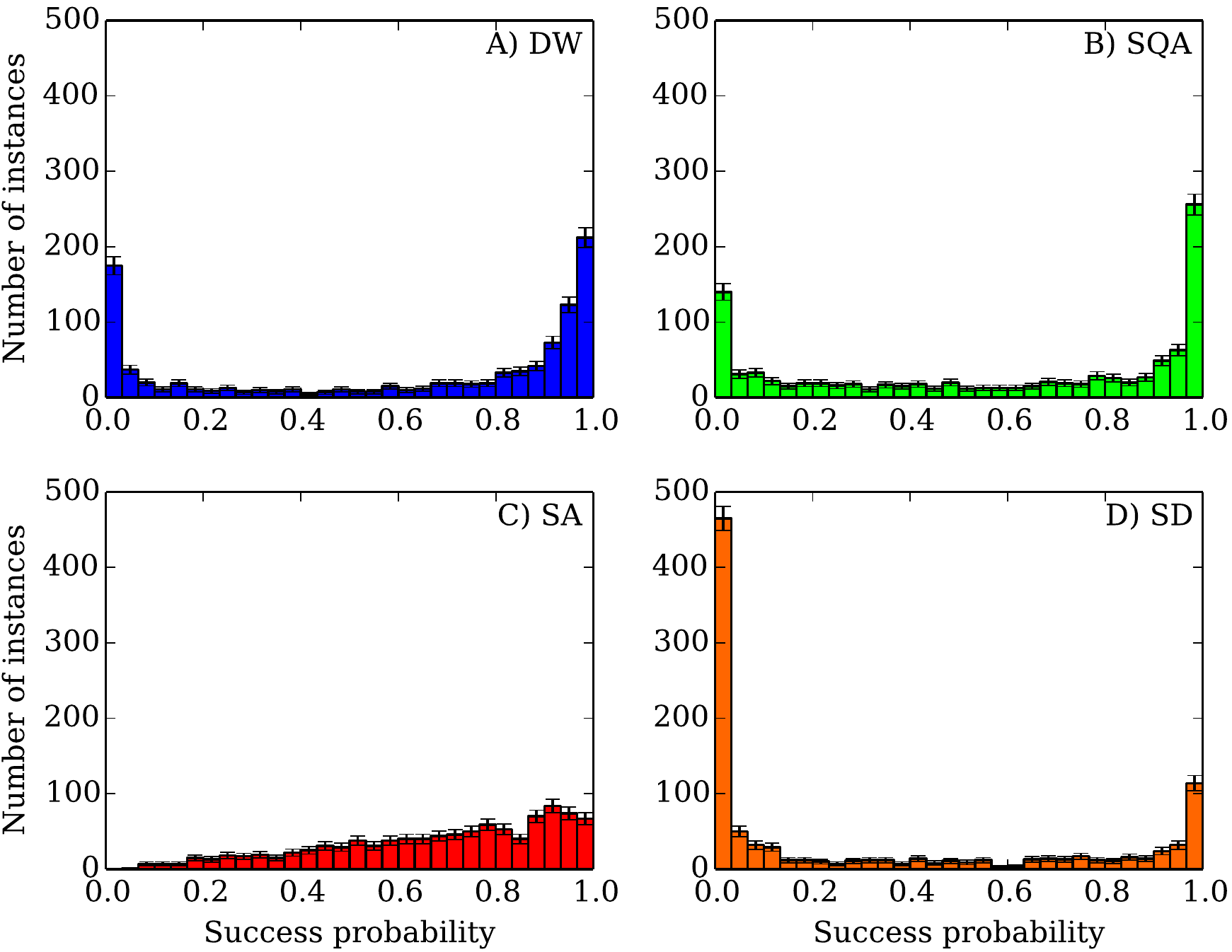}
\caption{Histogram of success probabilities from \cite{Boixo1}. It is observed that the histogram is bimodal for D-Wave, simulated quantum annealing, and classical spin dynamics, whereas it is unimodal for SA. This means that the former three algorithms divide the problem instances into two groups, namely ``easy'' and ``hard.'' They succeed almost always on the ``easy'' instances and fail almost always on the ``hard'' instances.}\label{fig:boixo_histogram}
\end{figure}
\begin{figure*}[!t]
\centering
\includegraphics[width=0.9\textwidth]{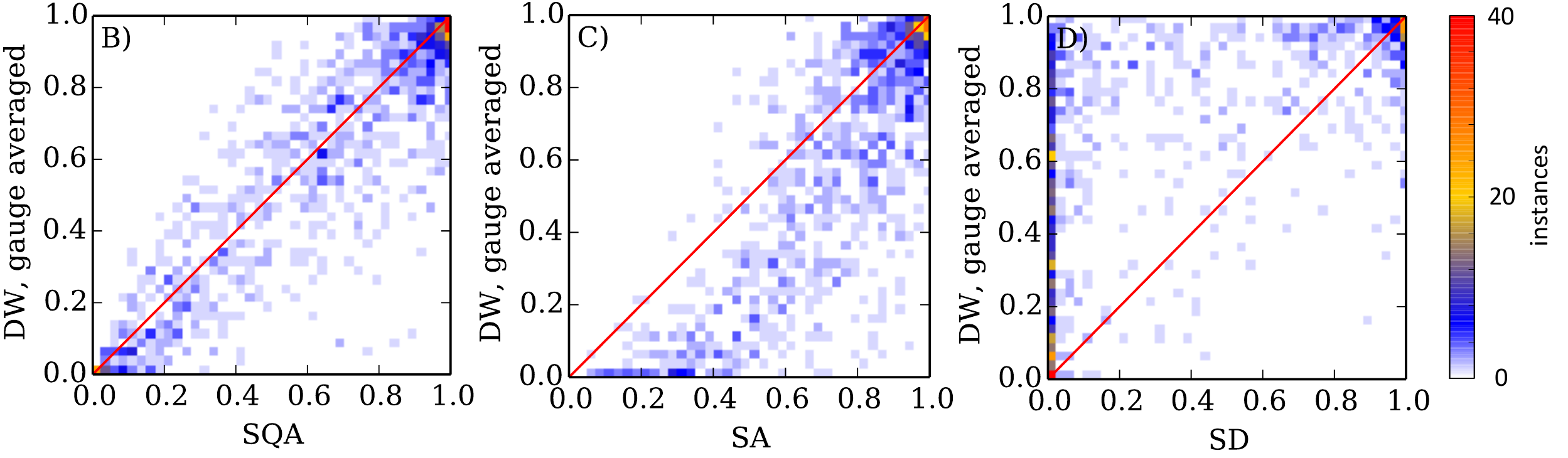}
\caption{Scatterplot of success probabilities from \cite{Boixo1}. The correlation between D-Wave and SQA is noticeably better than that between D-Wave and the classical models.}\label{fig:boixo_scatter}
\end{figure*}
\section*{Our model}

In our model, each spin $i$ is modeled by a classical magnet pointing in some direction $\theta_i$ 
in the XZ plane (since there is no $\sigma^y$ term in our time-dependent Hamiltonian, we assume that there is no $y$-component). We further assume that there is an external magnetic field of intensity $A(t)$ pointing in the $\hat{x}$-direction, and that neighboring magnets are coupled via either ferromagnetic or anti-ferromagnetic coupling, according to whether $J_{ij}$ is $1$ or $-1$. The resulting Hamiltonian mirrors the quantum Hamiltonian described earlier:
\[
H(t) = -A(t)\sum_{1\leq i\leq n} \sin \theta_i -B(t)\sum_{i<j} J_{ij} \cos \theta_i \cos \theta_j.
\]
Note that in the absence of noise, i.e.~at zero temperature, each spin $i$ will simply align with the net effective field at that location, $A(t)\hat{x} +B(t)\hat{z}\sum_{j} J_{ij} \cos \theta_j$. i.e. it makes an angle $\theta_i$ with the $z$-axis where $\tan{\theta_i} = A(t)/(B(t)\sum_{j} J_{ij} \cos \theta_j)$.

To simulate the effects of finite temperature $T$ of the system, we perform a Metropolis-type update. That is, at each time step, we pick a random angle $\theta_i'\in[0,2\pi)$ for each spin $i$ and update $\theta_i$ to $\theta_i'$ with probability $\max\{1,e^{-\Delta E_i/T}\}$ where
\begin{align*}
\Delta E_i=& -A(t)(\sin \theta_i'-\sin\theta_i) \\
&-B(t)\sum_{j} J_{ij} \cos \theta_j (\cos \theta_i'-\cos \theta_i).
\end{align*}

Note that our model is related to the $O(2)$ rotor model suggested in \cite{Smolin}, but differs in that we employ a more realistic noise process. Our update procedure can be regarded as the direct analogue of the Metropolis algorithm in $O(2)$ model.

\section*{Results}

To evaluate our model, we simulate it on the experimental data reported in~ \cite{Boixo1}, which presented the main evidence in favor of the existence of large-scale quantum effects in the D-Wave machines. That paper recorded the input-output behavior of D-Wave One on a thousand randomly chosen inputs, noting its probability of success in finding the exact ground state for each instance. It then compared this success probability to those of three different models: simulated quantum annealing,
(classical) simulated annealing and classical spin dynamics suggested in \cite{Smolin}. The paper produces two pieces of evidence; firstly, they observe that the histogram of success probabilities of D-Wave One is consistent with that of simulated quantum annealing rather than those of the two classical models (Figure \ref{fig:boixo_histogram}). Secondly, the correlation between the D-Wave success probabilities and SQA success probabilities is much higher than the correlation between the D-Wave success probabilities and success probabilities of the two classical models (Figure \ref{fig:boixo_scatter}).

\begin{figure}[!t]
\centering
\includegraphics[width=0.7\columnwidth]{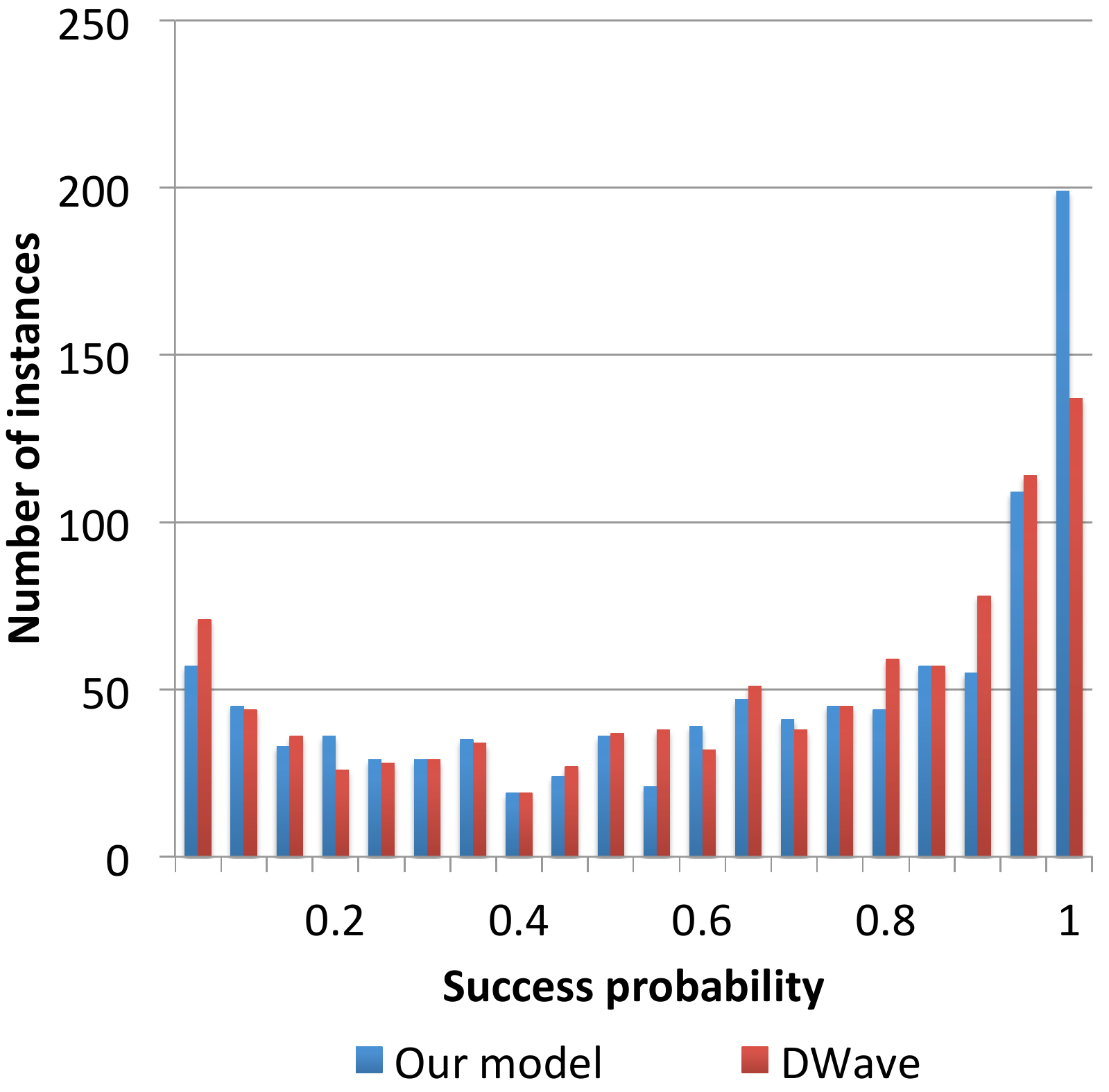}
\includegraphics[width=0.7\columnwidth]{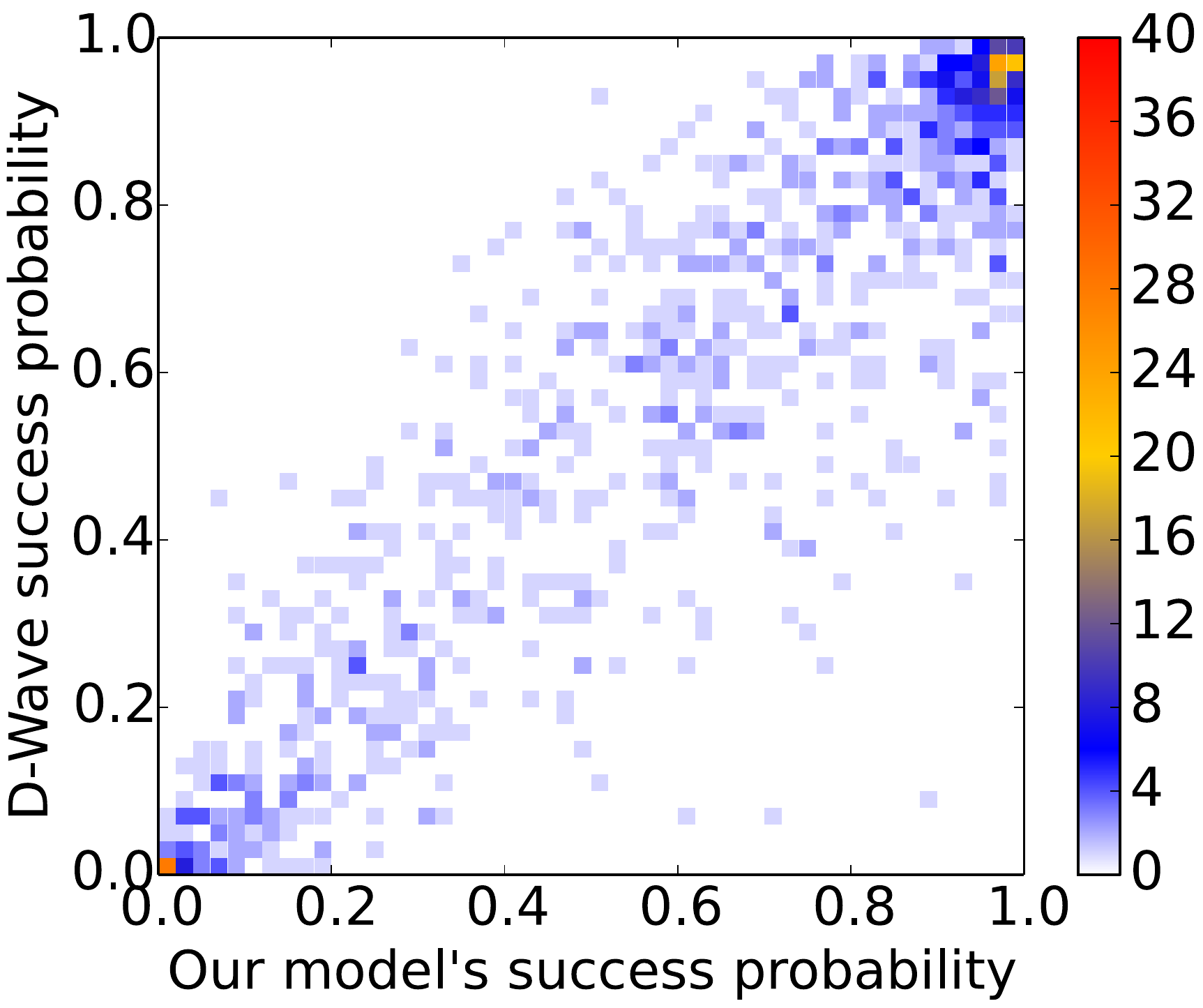}
\caption{Histogram and scatterplot of our classical model. Each run consisted of 150,000 steps and the system temperature of $T=0.22\text{GHz}\approx 0.9\text{mK}$ was used. The correlation coefficient $R$ between the D-Wave One and our model is about $0.91$.}\label{fig:result}
\end{figure}

Our experimental results on the same set of instances (Figure \ref{fig:result}) show that our simple classical model not only yields a histogram with clear bimodal signature similar to that of D-Wave One or simulated quantum annealing, but it also achieves a high correlation with the success probabilities of D-Wave One. 
 We note that the correlation of $0.91$ achieved by our model is slightly higher than the correlation achieved by simulated quantum annealing in \cite{Boixo1}.

\begin{figure}[!t]
\centering
\includegraphics[width=0.8\columnwidth]{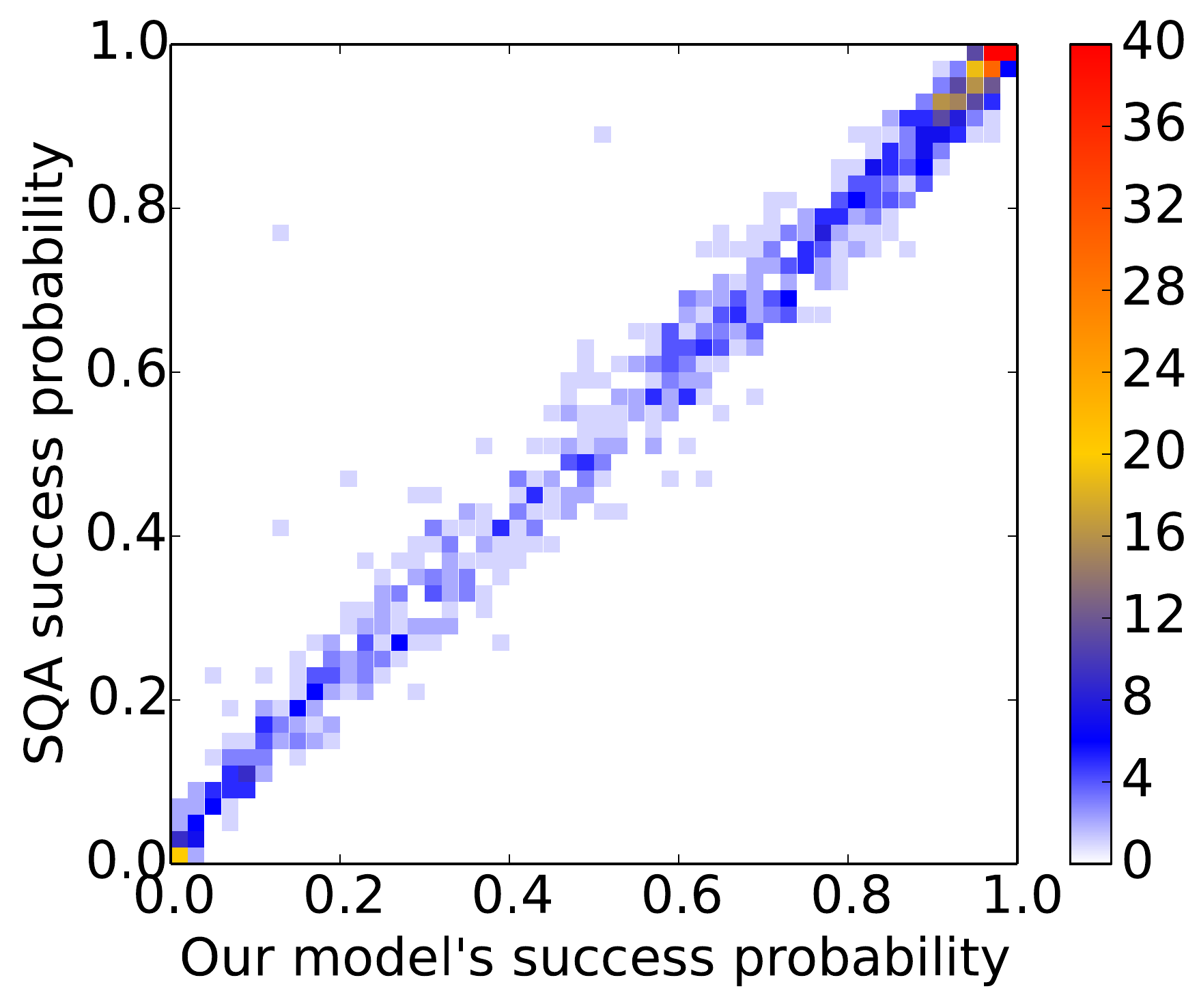}
\caption{Correlation between simulated quantum annealing of \cite{Boixo1} and our classical model. The correlation coefficient $R$ is about $0.99$.}\label{fig:sqacorrel}
\end{figure}

Matthias Troyer suggested~\cite{Troyer} a direct comparison between our model and simulated quantum annealing, which reveals an extremely high correlation of $R\approx 0.99$ (Fig~\ref{fig:sqacorrel}). In other words, our model can also be viewed as a mean-field approximation for quantum annealing where the system is assumed to be in a product state at every time step.
 
\section*{Discussion}

It is instructive to compare and contrast the behavior of our model with that of simulated annealing. We start by observing that our model simplifies to an $O(2)$ analogue of simulated annealing if $A(t)$ is set to be zero. This is because for the Metropolis acceptance probability function $e^{-\Delta E_i/T}$, keeping the temperature $T$ constant and increasing the coupling strength $B(t)$ over time has the same effect as keeping $B(t)$ constant and decreasing the temperature $T$ over time. 
Indeed, if we were to replace our model with $O(2)$ simulated annealing once $A(t)$ becomes sufficiently small (say after $t=0.31$), this does not change our experimental results. Moreover, since the effective temperature $T/B(t)$ is already small by this time, this is the regime in which simulated annealing behaves very much like greedy local search. 

The difference between our model and simulated annealing lies in the regime where the transverse field $A(t)$ is large (say when $t<0.31$). In simulated annealing, this is the part of the schedule where the system chooses randomly between a large number of low-energy local minima of the problem Hamiltonian. Note that this random choice explains why simulated annealing yields a unimodal histogram. By contrast, in our model, the time-dependent Hamiltonian admits only a very small number of local minima when $t$ is small. For example, it is easy to prove that it has only one local minimum when $t<0.06$, and it is empirically observed that our model reaches only a very small number of distinct local minima at $t=0.31$. Combined with the previous observation that in the second part of the schedule the algorithm is effectively greedy local search, we see that our model gives rise to more or less deterministic behavior in both the first and second parts of the schedule, which explains why it produces the bimodal histogram.

\begin{figure}[!t]
\centering
\includegraphics[width=.8\columnwidth]{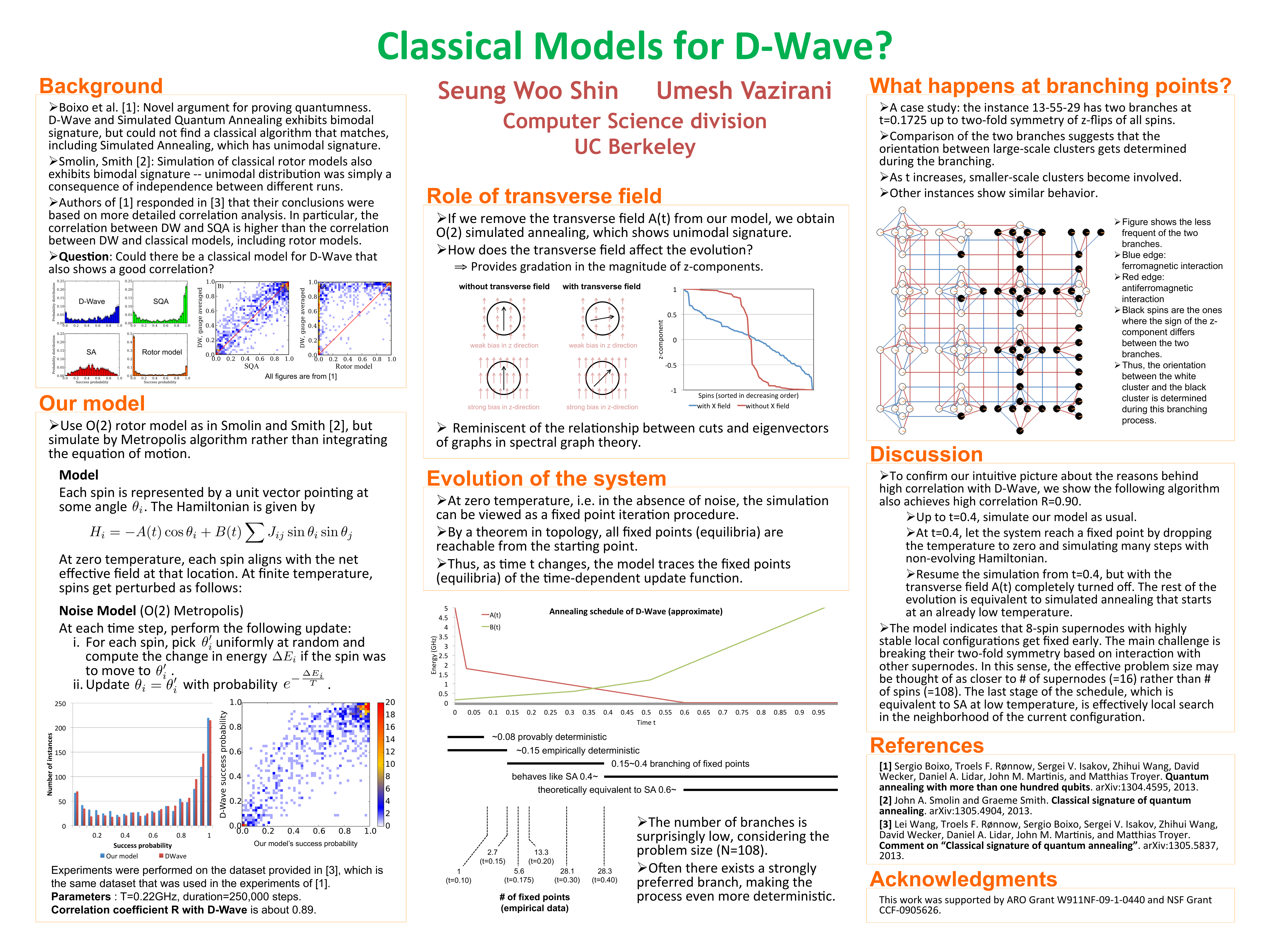}
\caption{Role of transverse field}\label{fig:transversefield}
\end{figure}

\begin{figure}[!t]
\centering
\includegraphics[width=.8\columnwidth]{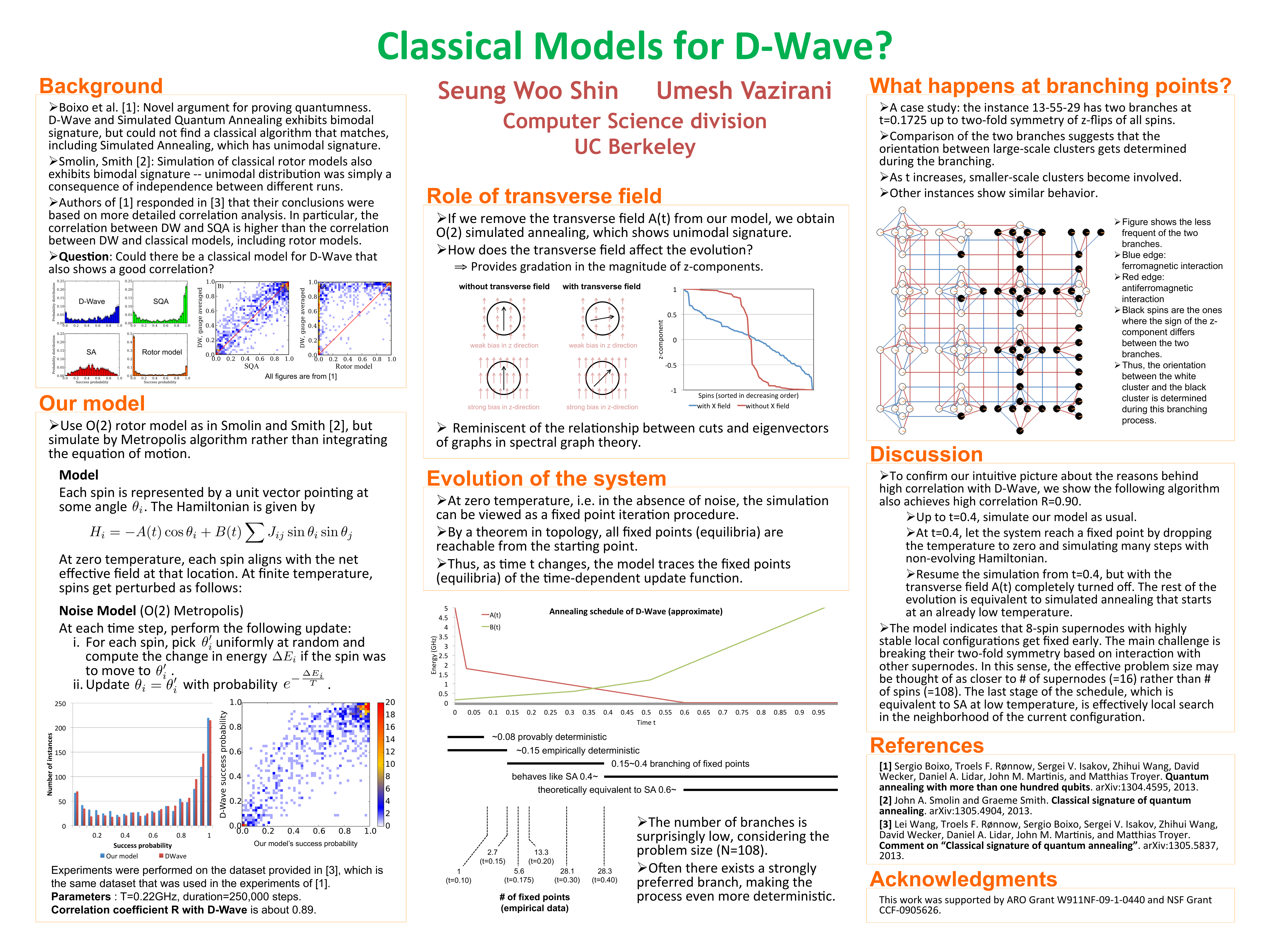}
\caption{Snapshot of a typical run: $z$-components of the spins, sorted in decreasing order.}\label{fig:cuteigenvector}
\end{figure}

The explanation for this difference in behavior lies in the gradation in the magnitude of $z$-components of spins in the presence of the transverse field (see Figures \ref{fig:transversefield} and \ref{fig:cuteigenvector}). The point is that in absence of a transverse field each spin will simply tend to point completely up or completely down, depending on the sign of the $z$ field at that location. In the presence of a transverse field the magnitude of the $z$ component determines the angle at which the spin is oriented. To draw an analogy with spectral graph algorithms, the two situations are like cuts versus eigenvectors of a graph. We also note that similar ideas are used in belief propagation algorithms.

\vspace{.2in}
\noindent
{\bf Clusters and Effective Problem Size}
\smallskip

\noindent Finally, we point out that averaged over all test instances, our simulations reached only about 20 distinct local minima at $t=0.31$. This is a surprisingly small number, considering that the state space is of size $2^{108}$ and that the model succeeds in solving a large fraction of the test instances. How is it that the model is able to solve such large problems by considering such a small number of possibilities? 

The answer lies in the structure of the ``Chimera" interaction graph, which consists of 16 clusters of supernodes of 8 vertices each: the density of edges within a supernode is much higher than the density of edges between supernodes. This makes it likely that many supernodes have highly stable configurations determined mostly by interactions across their internal edges. The two-fold symmetry consisting of flipping all the spins means that stable configurations come in pairs. To a first approximation, the major challenge for the algorithm consists of breaking this two-fold symmetry based on the energy contribution from interactions with other supernodes. 

\begin{figure}[!t]
\centering
\includegraphics[width=\columnwidth]{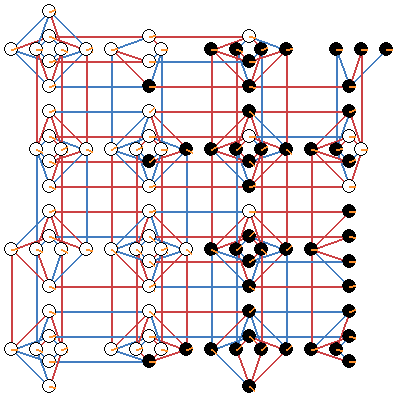}
\caption{The first ``branching point'' of instance 13-55-29 at $t=0.13$. Black dots indicate spins for which the signs of $z$-components differ between the two local minima. Figure shows the less frequent of the two local minima. Blue edge indicates ferromagnetic interaction and red edge antiferromagnetic interaction.}\label{fig:branching}
\end{figure}

A case study on the instance 13-55-29 illustrates this phenomenon very clearly. Our simulations reach two distinct local minima at $t=0.13$, up to the two-fold symmetry of $z$-flips of all spins. Figure \ref{fig:branching} shows that the choice being made at this ``branching point''  between these two local minima is the relative orientation between the ``white'' cluster and the ``black'' cluster. Note that the selection of the wrong branch will almost certainly cause the algorithm to fail, because the second part of the process, which we saw to be greedy local search, will not be able to flip such large clusters. An examination of other instances and values of $t$ strengthens this observation: branching points invariably correspond to points at which the orientation between large clusters gets determined. Moreover, as $t$ increases, we see smaller and smaller clusters become involved in the process. That is, we eventually see branching points which determine the orientation between one or two supernodes and the rest of the spins.

This observation provides a sense in which the effective problem size of Chimera-structured Ising ground state problem may be thought of as closer to the number of supernodes $m=16$ rather than the number of spins $n=108$. This analysis helps explain the success of the D-Wave machine on these problems despite their large apparent size. It also predicts that as the number of supernodes $m$ is scaled to be much larger than $16$, the success rate of the machine should fall dramatically. We remark that this conjecture seems to be supported by experimental evidence presented in a recent study \cite{RWJ14}. 

\section*{Acknowledgments}
SWS and UV were supported by ARO Grant W911NF-09-1-0440 and NSF Grant CCF-0905626. We thank Matthias Troyer, Sergio Boixo, and Troels F.~R{\o}nnow for useful discussions and for sharing their unpublished experimental data with us.

\bibliographystyle{naturemag}
\bibliography{dwave}

\end{document}